\documentstyle[twoside,fleqn,espcrc2]{article}

\def\be{\begin{equation}}
\def\ee{\end{equation}}
\input{psfig}

\newcommand{\AmS}{{\protect\the\textfont2
  A\kern-.1667em\lower.5ex\hbox{M}\kern-.125emS}}

\title{Universal correlations in spectra of the lattice QCD Dirac operator}

\author{M.A. Halasz\address{Dept. of Physics, SUNY at Stony Brook, 
        Stony Brook, NY\,11794},
        T. Kalkreuter\address{IS Internet Services, D-21079, Hamburg, Germany}
        and
        J.J.M. Verbaarschot$^{ \rm a}$}
       
\begin{document}

\begin{abstract}
Recently, Kalkreuter obtained complete Dirac spectra
for $SU(2)$ lattice gauge theory both for staggered fermions
and for Wilson fermions. The lattice size was as large as
$12^4$. We performed a statistical analysis of these data and found
that the eigenvalue correlations can be described by the Gaussian
Symplectic Ensemble for staggered fermions and by the Gaussian
Orthogonal Ensemble for Wilson fermions. In both cases
long range spectral fluctuations are strongly
suppressed: the variance of a sequence of levels containing
$n$ eigenvalues on average is given by
$\Sigma_2(n) \sim 2 (\log n)/\beta\pi^2 $ ($\beta$ is 
equal to 4 and 1, respectively)
instead of $\Sigma_2(n) = n$ for a random sequence of levels.
Our findings are in agreement with the anti-unitary symmetry of the lattice
Dirac operator for $N_c=2$ with staggered fermions which differs
from  Wilson fermions (with the continuum anti-unitary symmetry).
For $N_c = 3$,  we predict that the
eigenvalue correlations are given by the Gaussian Unitary Ensemble.
\end{abstract}

\maketitle

\section{INTRODUCTION}
The study of spectra of the QCD Dirac operator is of great importance
for the understanding of the mechanism of chiral symmetry breaking and its
restoration above the critical temperature $T_c$. Moreover, the fluctuations of 
the eigenvalues provide information on the distribution of the fermion 
determinants, and ultimately, on the validity of the quenched approximation.
 In this work, we study the Dirac spectrum for $SU(2)$ color both with  
Kogut-Susskind (KS) and Wilson fermions.

We consider the eigenvalue problems \cite{Kalkreuter}
\be
iD^{\rm KS} \psi = \lambda \psi,
\ee
\be
Q_5^{\rm W} \psi\equiv  \frac 1{8+m} \gamma_5 (D^{\rm W} + m) 
\psi = \lambda \psi,
\ee
where $D^{\rm KS}$ and $ D^{\rm W}$ are the usual KS and
Wilson lattice QCD Dirac operators.
The Dirac matrix is tri-diagonalized by Cullum's and Willoughby's Lanczos
procedure \cite{Cullum}. 
The eigenvalues are then obtained by a standard QL algorithm. 
This improved Lanczos algorithm makes it possible to obtain {\it all}
eigenvalues. The accuracy of the eigenvalues
can then be checked by means of sum rules for the sum of the squares
of the eigenvalues of the lattice Dirac operator \cite{Kalkreuter}.
\section{SPECTRA OF COMPLEX SYSTEMS AND RANDOM MATRIX THEORY}
A basic assumption in the analysis of spectra of complex systems is
that the variations of the average spectral density, $\bar \rho(\lambda)$,
and the spectral 
fluctuations about this average separate. This allows us to unfold 
the spectrum $\{\lambda_k\}$ into
$\{ \lambda_k'\}$, with average spectral 
density equal to one,  according to
$
\lambda_k' = \int_{-\infty}^{\lambda_k}d\lambda \bar \rho(\lambda).
$
We study the distribution of the number
of unfolded eigenvalues, $\{ n_k(n)\}$, in a stretch of length $n$,
by means of its variance, 
$\Sigma_2(n) $, and its  first two cumulants,
$\gamma_1(n)$ and $\gamma_2(n)$.

These statistics can be obtained analytically for the invariant random matrix
ensembles of hermitean matrices with 
probability distribution given by $P(H) \sim \exp(-{\rm Tr} \, H^\dagger H )$.
The matrix elements can be either real (Gaussian Orthogonal Ensemble (GOE)),
complex (Gaussian Unitary Ensemble (GUE)) or quaternion real (Gaussian 
Symplectic Ensemble (GSE)). They are characterized by the Dyson index $\beta$,
which is equal to 1, 2 and 4, respectively.
The most notable property of random matrix
ensembles is the stiffness of the spectrum, i.e. $\Sigma_2(n) \sim 
(2/\pi^2\beta) \log(n)$, instead of $n$ for independently distributed 
eigenvalues.

In the past decade, spectra of many complex systems have been studied in
great detail. The main conclusion \cite{univers} is that, 
if the corresponding classical
system is chaotic, the spectral correlations are given by the random 
matrix theory (RMT) with the same (anti-unitary) symmetries as the original 
Hamiltonian. We also note that the average spectral density is generically 
not given by RMT. 
\section{SPECTRAL CORRELATIONS OF THE DIRAC OPERATOR}
\begin{figure}[htb]
\psfig{figure=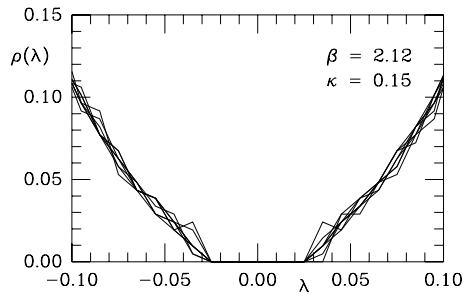,width=60mm}
\vspace*{-0.7cm}
\caption{A histogram of the spectral density of Wilson fermions for 8
independent configurations.}
\label{density}
\vspace*{-0.7cm}
\end{figure}

For a lattice size of $6^3\times 12$ and Wilson fermions with $\beta =2.12$
and $\kappa = 0.15$, all eigenvalues
were obtained for 8 independent gauge field configurations. 
A histogram of the spectral density (normalized to 1) 
in the region $[-0.1, 0.1]$ of each of
these configurations is shown in figure \ref{density}. The fluctuations of
the Dirac spectrum over
the ensemble and the statistical fluctuations from bin to bin seem to
coincide. This so called
'spectral ergodicity', the equality of ensemble averages and 
spectral averages, is a well known property
of RMT \cite{Pandey}.
We have also calculated the
ensemble average of the number variance of these spectra unfolded 
with the ensemble averaged spectral density. Within the limited statistics
we found complete agreement with the GOE prediction.
For long level sequences much better statistics can be obtained from a
spectral average rather than an ensemble average which we will use in
the remainder of the paper. Below we will show that
the spectral average of $\Sigma_2(n)$ for Wilson fermions 
is given by the GOE as well. 

\begin{figure}[htb]
\psfig{figure=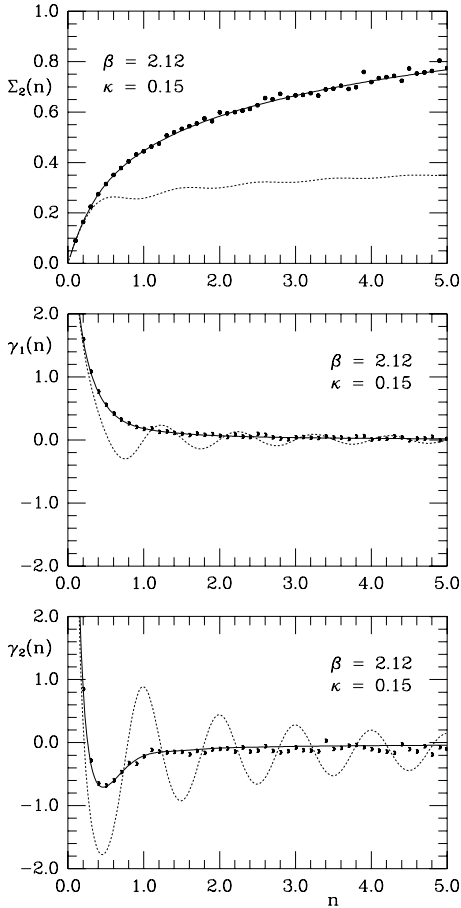,width=60mm}
\vspace*{-0.5cm}
\caption{Spectral correlations for Wilson fermions.}
\label{Wilson}
\vspace*{-0.5cm}
\end{figure}

\begin{figure}[htb]
\psfig{figure=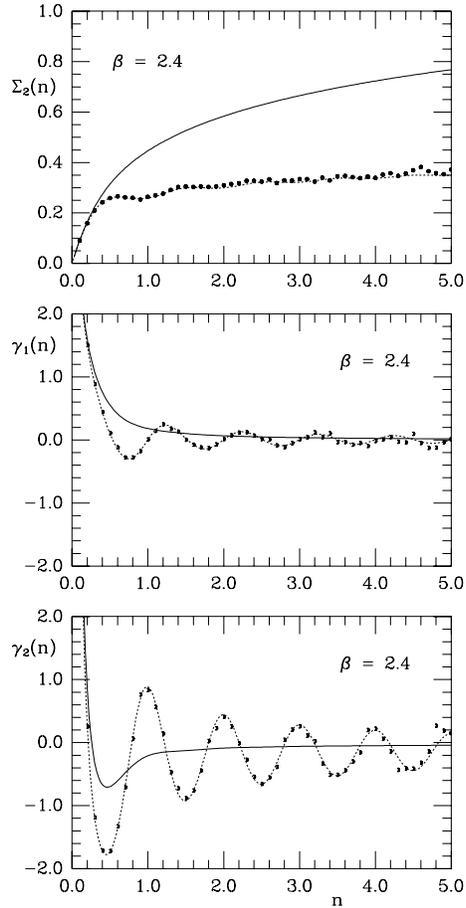,width=60mm}
\vspace*{-0.5cm}
\caption{Spectral correlations for KS fermions.}
\label{KS}
\vspace*{-0.5cm}
\end{figure}

For $SU(2)$, the anti-unitary symmetry of the KS Dirac operator
 and the Wilson Dirac operator is different. We have $[\tau_2 K, D^{\rm 
KS}] = 0$ and $[\gamma_5 C\tau_2 K, Q_5^{\rm W}] = 0$, where $K$ is the complex
conjugation operator and $C$ the charge conjugation matrix. 
Because $(\tau_2 K)^2 = -1$, whereas $(\gamma_5 C\tau_2 K)^2 = 1$ 
the KS Dirac matrix can be organized into
real quaternions, whereas the Wilson Dirac matrix is real in an
appropriate basis. Consequently, we expect that the spectral correlations
of the eigenvalues of $D^{\rm KS}$ are given by the GSE, and those of
$Q_5^{\rm W}$ are given by the GOE.
In Figs. \ref{Wilson} and \ref{KS} we show the number variance, $\Sigma_2(n)$,
and the first two cumulants, $\gamma_1(n)$ and $\gamma_2(n)$, versus $n$.
The points are obtained from the unfolded lattice spectra, with the 
average integrated spectral density obtained by fitting a second order
polynomial to a stretch of 500-1000 eigenvalues. Analytical results are 
represented by the full (GOE) and dotted curves (GSE).
The KS results are for a $12^4$ lattice with
4  dynamical flavors with a mass of $ma=0.05$. The results for dynamical
Wilson fermions were obtained on a $8^3\times 12$ lattice. The values of $\beta$ and $\kappa$
are shown in the label of the figure.
Clearly, perfect agreement with the random matrix prediction is observed.
We also studied spectral correlations for weaker coupling $\beta =2.8$, and
stronger coupling $\beta = 1.8$, and both quenched and unquenched field
configurations. The number variance was calculated up to $n = 100$ 
\cite{HV} 
The stationarity of the spectral correlations was investigated
by studying correlations of stretches of 200 eigenvalues starting from zero.
In all cases no deviations from the random matrix predictions
were found. 
\section{DISCUSSION AND CONCLUSIONS}
We have seen that the spectral correlations of the KS Dirac
operator are given by the GSE, contrary to those of the Wilson Dirac operator,
with the anti-unitary symmetries of the continuum theory 
which are given by the 
GOE. One might wonder what happens to the spectral correlations 
of $D^{KS}$ in the
continuum limit. In a suitable basis, we can
we write \cite{Golt} $D^{\rm KS} = D^{\rm cont. } + a^2 B$.
The operator $a^2 B$ is relevant for the spectral fluctuations if its
norm is larger than the level spacing of $D^{\rm cont. }$. 
At fixed volume,  the level spacing of $\lambda a$ 
in the bulk of the spectrum is of order 
$1/N= a^4/V$, which is much less than the norm of the
perturbing operator. Only for the low-lying spectrum with $\lambda
\sim 1/V$ for $a \rightarrow 0$, a transition to the GOE is possible.
Apparently, much larger lattices are required to study this transition.

Finally, for $SU(3)$ color, the Dirac operator does not have
any additional anti-unitary symmetries, and 
we predict that the spectral correlations for both
KS fermions and Wilson fermions are given by the GUE. 
In all cases the spectral fluctuations are strongly
suppressed with respect to uncorrelated eigenvalues. To some extent, QCD
is selfquenching.

The reported work was partially supported by the US DOE grant
DE-FG-88ER40388.


\begin{thebibliography}{9}
\bibitem{Kalkreuter}T. Kalkreuter,  Comp. Phys. Comm. {\bf 95} (1996) 1;
Phys. Lett. {\bf B276} (1992) 485; Phys. Rev. {\bf D48} (1993)
\bibitem{Cullum}J. Cullum and R.A. Willoughby, J. Comp. Phys. {\bf 44} (1981) 
329.
\bibitem{univers}O.~Bohigas, M.~Giannoni, Lecture notes in Physics
{\bf 209}, Springer Verlag 1984, p. 1.
\bibitem{HV}M.A. Halasz and J.J.M. Verbaarschot,
Phys. Rev. Lett. {\bf 74} (1995) 3920; J.J.M. Verbaarschot, this proceedings.
\bibitem{Pandey} A. Pandey, Ann. Phys. {\bf 134} (1981) 119.
\bibitem{Golt} M.F.L. Golterman and J. Smit, Nucl. Phys. {\bf B245} (1984) 61;
Y. Luo, hep-lat/9604025.

\end{thebibliography}
\end{document}